\documentclass[iop]{emulateapj}
\usepackage{amsmath}
\usepackage{amssymb}
\usepackage{latexsym}
\usepackage{longtable}
\usepackage{verbatim}
\usepackage{natbib}
\usepackage{mathtools}
\usepackage{multirow}
\usepackage{empheq}
\usepackage{bm}
\usepackage{graphicx}
\usepackage{color}

\newcommand{\Ha}{H$\alpha$\,$\lambda$6563}
\newcommand{\Hb}{H$\beta$\,$\lambda$4861}

\newcommand{\OIII}{[O\,{\sc iii}]\,$\lambda$5007}

\newcommand{\LOIII}{$L_{\rm [O\,{\scriptsize \textsc{iii}}]}$}
\newcommand{\OIIII}{{\rm [O\,{\sc iii}]\,$\lambda$5007}}
\newcommand{\NII}{[N\,{\sc ii}]\,$\lambda$6584}

\begin{document}
\title{Stellar Populations of a Sample of Optically Selected AGN-host Dwarf Galaxies}
\author
{Wei Cai\altaffilmark{1,2,3},
Yinghe Zhao\altaffilmark{1,3,4,*},
Hong-Xin Zhang\altaffilmark{5},
Jin-Ming Bai\altaffilmark{1,3,4},
and
Hong-Tao Liu\altaffilmark{1,3,4}}
\altaffiltext{1}
{Yunnan Observatories, Chinese Academy of Sciences, Kunming 652016, China}

\altaffiltext{2}
{University of Chinese Academy of Sciences, Beijing 100049, China}

\altaffiltext{3}
{Key Laboratory for the Structure and Evolution of Celestial Objects, Chinese Academy of Sciences, Kunming 650216, China}

\altaffiltext{4}
{Center for Astronomical Mega-Science, Chinese Academy of Sciences, Beijing 100012, China}

\altaffiltext{5}
{CAS Key Laboratory for Research in Galaxies and Cosmology, Department of Astronomy, University of Science and Technology of China, Hefei, Anhui 230026, China}

\altaffiltext{*}{Corresponding author: YZ (zhaoyinghe@ynao.ac.cn).}
\begin{abstract}
In this paper we present our studies on the stellar populations and star formation histories (SFHs) for the Reines et al. sample of 136 dwarf galaxies which host active galactic nuclei (AGNs), selected from the Sloan Digital Sky Survey Data Release 8. We derive stellar populations and reconstruct SFHs for these AGN-host dwarfs using the stellar population synthesis code STARLIGHT. Our results suggest that these AGN-host dwarfs have assembled their stellar masses within a narrow period of time with the stellar mass-weighted ages in the range of $10^9-10^{10}$~yr, but show a wide diversity of SFHs with the luminosity-weighted stellar ages in the range of $10^7-10^{10}$~yr. The old population ($t>10^9$~yr) contributes most to the galaxy light for the majority of the sample; the young population ($t<10^8$~yr) also appears in significant but widely varying fractions, while the intermediate-age population ($10^8<t<10^9$~yr) in general contributes less to the optical continuum at 4020 \AA. We also find that these dwarfs follow a similar mass-metallicity relation to normal star-forming galaxies, indicating that AGNs have little effect on the chemical evolution of the host galaxy. We further investigate the relation between the derived SFHs and morphology of the host galaxy, and find no correlation. Comparing the SFHs with the luminosity of the \OIII\ line ($L_{\rm [O\,{\scriptsize \textsc{iii}}]}$), we find that there exists a mild correlation when $L_{\rm [O\,{\scriptsize \textsc{iii}}]} > 10^{39}$~erg~s$^{-1}$, indicating that there is a physical connection between star formation and AGN activities in these dwarf galaxies.
\end{abstract}

\keywords{galaxies: dwarf - galaxies: active - galaxies: statistics - galaxies: stellar content - galaxies: nuclei.}

\section{INTRODUCTION}
The feedback of AGNs on host galaxies has been a fundamental component in cosmological simulations, and it is widely accepted that AGNs have significant effects on the evolution of massive galaxies (see the reviews by Kormendy \& Ho 2013 and Heckman \& Best 2014). AGNs may be triggered by the gas inflows into the nuclear or mass-loss from evolving stars (e.g., Wild, Heckman \& Charlot 2010). Meanwhile, the gas may also be used to form stars. Therefore, the stellar populations of AGN host galaxies are an important tool for studying the connection between AGN and star formation. A series of works have tried to derive the stellar populations of AGN-host massive galaxies by using the nuclear spectra (e.g., Cid Fernandes et al. 2004; hereafter C04) and found that AGN activities might be related to recent episodes of star formation in the nuclear region (e.g., Storchi-Bergmann et al. 2001; Rembold et al. 2017). There also exist a few works studying the spatially resolved stellar populations of AGN host galaxies (e.g., S{\'a}nchez et al.  2018; Mallmann et al. 2018). Some works also tried to correlate nuclear star formation histories (SFHs) with other properties such as AGN luminosity (e.g., Rembold et al. 2017) and morphology of AGN host galaxies (e.g., Storchi-Bergmann et al. 2001; C04; Mallmann et al. 2018).

AGNs have been found in dwarf galaxies (e.g., NGC 4395; Filippenko \& Sargent 1989; POX 52; Barth et al. 2004), which are the most numerous galaxies in the Universe (e.g., Mateo 1998), and have been proposed to be building blocks from which more massive systems are formed via merging (Kauffmann et al. 1993). Dwarfs are typically metal-poor (e.g., Mateo 1998), making them having similar conditions of high-redshift galaxies. Furthermore, the first seed black holes may have masses not so different from those in dwarf galaxies due to their relatively quiet merger histories (Bellovary et al. 2011). Therefore, dwarf galaxies are a good place to explore the connection between AGN activities and star formation.

However, little attention has been paid to study the role that AGNs play in the evolution of dwarf galaxies until recent years, which might be due to the fact that there are very limited number of AGNs identified in dwarf galaxies before large spectroscopic surveys (such as the Sloan Digital Sky Survey, SDSS; York et al. 2000) were conducted.  Furthermore, the importance of the AGN effects on the evolution of dwarfs is still in debate. Both the theoretical model in Dashyan et al. (2018) and the cosmological simulation in Barai \& de Gouveia Dal Pino (2019) suggest that AGN feedback can effectively quench the star formation of host galaxies, while other simulations give opposite results that AGNs have ignorable effects on the star formation of dwarf galaxies (e.g., Trebitsch et al. 2018; Koudmani et al. 2019). Observationally, Penny et al. (2018) show some evidence that  AGNs inhibited star-formation in several low-mass galaxies ($M_{\star}<5 \times10^{9}M_{\odot}$), and Manzano-King et al. (2019) find that the outflows in six out of nine AGN-host dwarf galaxies are mainly driven by the central AGN. For a sample of AGN-host dwarf galaxies at $z\sim3.4$, Mezcua et al. (2019) show that the radio jets can be as powerful as those of massive radio galaxies, and thus AGN feedback may also have a very strong impact on the host dwarf galaxies.

The stellar populations of the host galaxy are expected to be related to AGN effects and thus there should exist some difference between normal and AGN-host dwarf galaxies if the latter have been significantly affected by AGN activities. For normal dwarfs, there have been many studies on the stellar populations (e.g., Tolstoy et al. 2009; McQuinn et al. 2010; Weisz et al. 2011a,b; Zhao et al. 2011; Kauffmann 2014; Rowlands et al. 2018), using both photometric and (integrated and spatially-resolved) spectroscopic data. Based on optical color-magnitude diagrams, Weisz et al. (2011a,b) show that the mass assembly histories of typical dwarfs in the Local Group and Local Volume are dependent on morphological types, with 90\% (dwarf irregular: dI) to 99\% (dwarf spheroidals/ellipticals: dSph/dE) of their stellar mass formed more than 1~Gyr ago. Using integrated spectra and a much larger sample, Kauffmann (2014) further demonstrates that the fraction of present-day stellar mass ($M_\star$) contributed by young stellar populations ($t<10^8$ yr) increases as $M_\star$ decreases.  

However, few works have studied the stellar populations in AGN-host dwarfs. In this paper, we present a detailed study of the stellar populations for hitherto the largest sample of AGN-host dwarf galaxies containing 136 sources, using a simple stellar population (SSP) synthesis method. This is capable of yielding the various stellar components, SFHs, AGN contribution to the optical continuum and internal extinction. We also investigate the dependence of the nuclear stellar populations on other galactic properties such as morphology and \OIII\ luminosity. Combining with literature results of normal dwarfs, our studies might shed some light on the effects of AGNs on their hosts.
 
 The paper is organized as follows. Section 2 describes the AGN sample and our data reductions. Our results and analysis are given in Section 3. We discuss the aperture effects of the fiber spectra, the effect of the AGN featureless continuum on estimation of stellar population properties in Section 4 and summarize our results in the last section. Where required we adopt a Hubble constant of $H_{0}$ = 73 km s$^{-1}$ Mpc$^{-1}$, the same as Reines et al. (2013; hereafter R13).

\section{SAMPLE, DATA, AND METHOD}
\subsection{Sample and Data}
 Here we adopt the AGN host dwarf galaxy sample from R13, who have used the 2D line-intensity ratio calculated from relatively strong lines of \Hb, \OIII, \Ha\ and \NII\ to probe the nebular conditions of a source (i.e. BPT diagram; Baldwin, Phillips \& Terlevich 1981; Veilleux \& Osterbrock 1987). In brief, R13 searched active BHs from a large sample (25,974 members) of emission-line galaxies with stellar masses comparable to the Magellanic Clouds ($M_{\star}\leqslant3\times10^{9}$$M_{\odot}$) and redshifts $z<0.055$, using the data from the Sloan Digital Sky Survey Data Release 8 (SDSS DR8; Aihara et al. 2011) with stellar masses adopted from the NASA-Sloan Atlas\footnote{http://www.nsatlas.org}(NSA). They found that 136 dwarf galaxies have optical signatures of  AGN activity in the BPT diagram, out of which 35 have AGN dominated spectra and 101 have composite spectra. The sample galaxies have a median redshift of $\langle z \rangle = 0.028$, a median $g$-band absolute magnitude of $\langle M_g \rangle \sim -18.1$ mag, a median $g-r$ color of $\langle g-r \rangle =0.51$ mag, and a median stellar mass of $\langle M_\star \rangle =2\times10^9\,M_\odot$. The physical size (diameter) covered by the SDSS 3\arcsec-fiber is in the range of 12 pc to 3.3 kpc, with a median value of $\sim$1.8~kpc.

The current sample is not complete in any sense, as it is difficult to achieve a complete sample for faint galaxies. Situations become even worse for the selection of AGN host dwarf galaxies since a weak AGN signature might be overwhelmed by the light from its host galaxy (e.g., Trump et al. 2015). Bearing this caveat in mind, however, our sample galaxies are selected from a homogeneous dataset. It is the largest sample of dwarfs with optical signature of AGNs to date, and spans a large range of galactic parameters including stellar mass ($M_\star$ of 10$^{8.1}$$-$$ 10^{9.5}$~$M_\odot$), integrated magnitude ($M_g$ of $-15$ to $-21.5$ mag), color ($g-r$ of $-0.3$$-$1.0 mag), size (the Petrosian 50\% light radius $r_{50}$ of 0.05$-$7~kpc) and morphology (S{\'{e}rsic index $n$ of 0.5$-$6), which cover similar ranges of respective parameters of the parent sample (see Figure 7 in R13 for details), except for that there are few AGNs found in the faint and/or blue end. The redder color of this AGN sample is likely caused by a selection effect because the optical diagnostics is not sensitive to AGNs when the star-forming host galaxies dominates the emission-line spectra.

 Furthermore, in order to derive the SFHs of this AGN-host dwarf sample, we need to check to what extent the SDSS spectrum represents the light of the entire galaxy. This is  because  the SDSS is a fiber-based survey, and the spectrum only covers the starlight from a $3\arcsec$-diameter aperture. For our sample galaxies, the minimum physical size covered by the SDSS fiber is $\sim$12~pc for the nearest object (SDSS J122548.86+333248.7), and the maximum size is $\sim$3.3~kpc for the most distant object (SDSS J101747.09+393207.7). We further adopted the light fraction ($f_r$) covered by the fiber in the $r$-band to describe the covering fraction of the fixed-size aperture for the AGN sample, as in Zhao et al. (2011). To calculate $f_r$, we used the fiber magnitude and the total galaxy magnitude, explicitly $f_r = 100 \times 10^{-0.4(m_{\rm fiber}-m_{\rm Petro})}$. As shown in Figure \ref{figfr}, the median value of $f_r$ is 36.3\%, which is about twice larger than the value (20\%) required to minimize the aperture effects in the spectral measurements (Kewley et al. 2005). In the following analysis, we further divide our sample galaxies into two subsamples according to this threshold: $f_r < 20\%$ and $f_r \geq 20\%$, and we refer to them as S1 and S2, respectively. There are 37 (99) galaxies in S1 (S2), and the median $f_r$ of S1 (S2) is $\sim$8\% ($\sim$50\%). Therefore, sources in S1 might severely suffer from the aperture effect. We will discuss how our conclusions would be affected by the aperture effect later.

 \begin{figure}[t]
\centering
          \includegraphics[width=0.47\textwidth,bb=35 38 275 275]{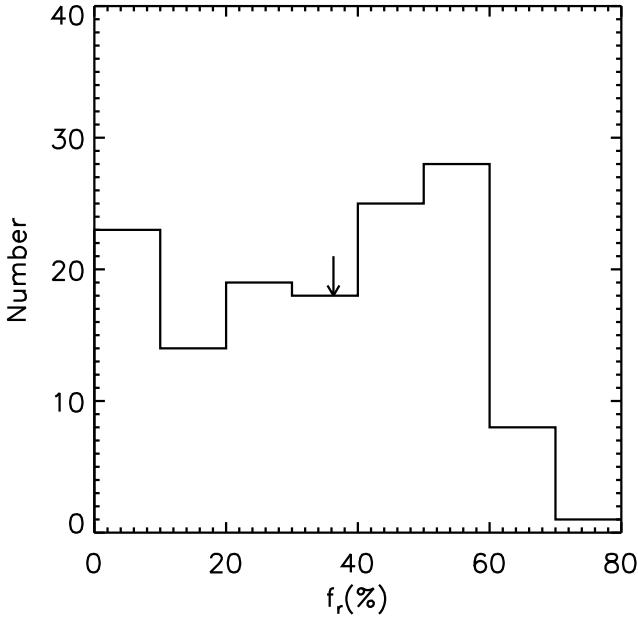}
\caption{The number distribution of $f_r$ for the AGN host dwarf sample. The arrow shows the median value of $f_r$ (36.3\%).  There are 37 and 99 galaxies  with $f_r < 20\%$  and  $f_r > 20\% $, respectively.}
\label{figfr}
\end{figure}   
    
\subsection{Stellar Population Synthesis}
\begin{figure*}[t!]
\centering
	\includegraphics[width=0.95\textwidth,bb=20 110 580 457]{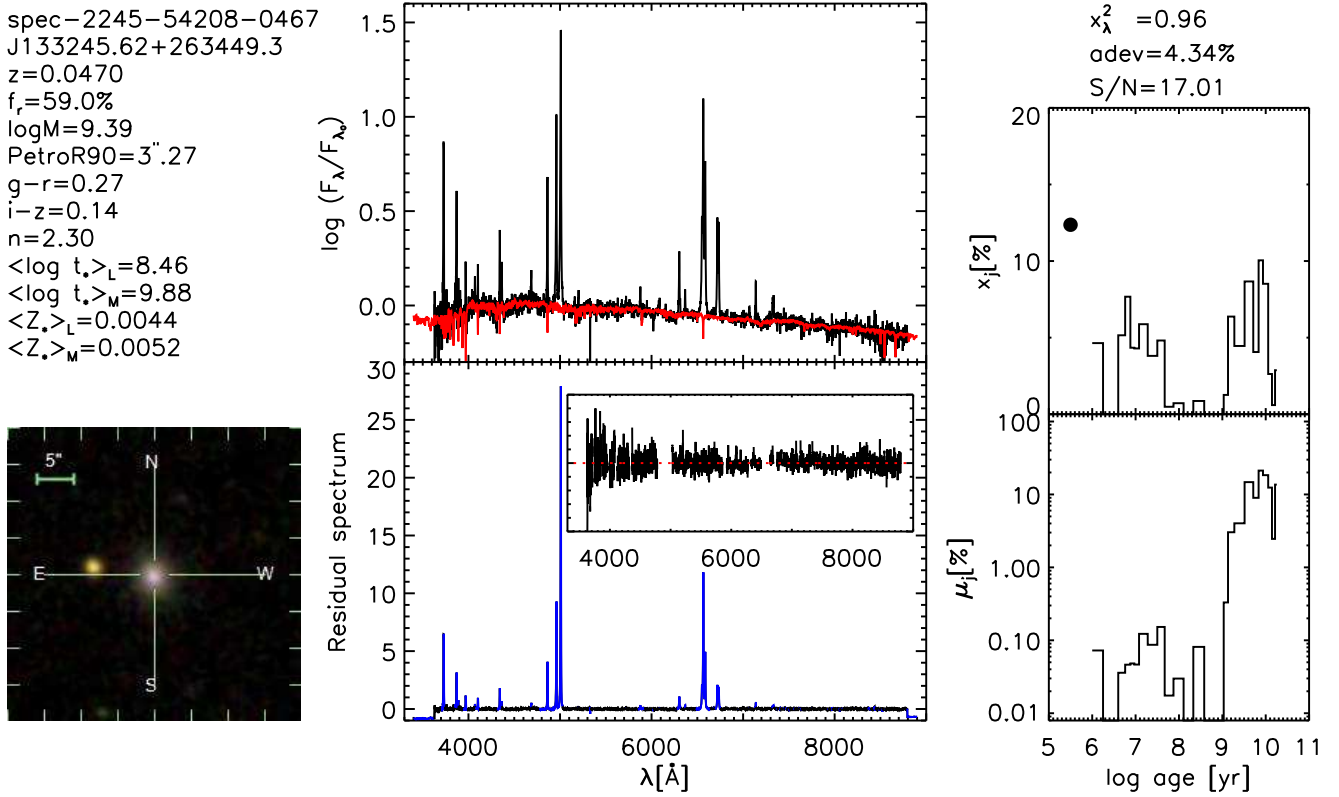}
\caption{Results of the spectral fitting for SDSS J133245.62+263449.3.  The top middle panel shows the logarithm of the observed ($F_{\lambda}^O$ ; black line) and the synthetic $F_{\lambda}^M$; red line) spectra, normalized to $\lambda_0\sim4020,\r{A}$. The $F_{\lambda}^O-F_{\lambda}^M$ residual spectrum is shown in the bottom middle panel.  The inset illustrates the zoomed-in view of the continuum, with the red dashed line showing the constant value of 0. Spectral regions actually used in the synthesis are plotted with a black line, while masked regions replotted with a blue line. Panels in the right show the population vector binned in the 25 ages of SSPs used in the model library. The top right panel corresponds to the population vector in flux fraction, normalized to $\lambda_0 = 4020,\r{A}$, while the corresponding mass fraction vector is shown in the bottom right panel. The power-law component $x_{\rm AGN}$ is plotted with an (arbitrary) age of $10^{5.5}$ yr and marked by a black point.  The left panel gives the three color-RGB image (R: $i$-band; G: $r$-band; B: $g$-band; Lupton et al. 2004) obtained from the SDSS Science Archive Server, as well as the information of this object.}
\label{figexample}
\end{figure*}

In the current work, we adopt the same process as that in Zhao et al. (2011) to model the stellar contribution in the SDSS spectra for the AGN-host dwarfs. In brief, the stellar population synthesis code, STARLIGHT (Cid Fernandes et al. 2005, hereafter C05, 2007; Mateus et al. 2006; Asari et al. 2007), is used to fit the observed spectrum with a linear combination of $N_{\star}$ simple stellar populations (SSPs). Here we adopt a base with $N_{\star}=100$ (25 ages from 1 Myr to 18 Gyr, and 4 metallicities of Z=0.005, 0.02, 0.2, and 0.4$Z_{\odot}$), which were computed with the Salpeter (1955) initial mass function, Padova-1994 models, and the STELIB library (Le Borgne et al. 2003) using the evolutionary synthesis models published by Bruzual \& Charlot (2003, hereafter BC03). The fractional contribution of each SSP to the total synthetic flux at the  normalization  wavelength $\lambda_0$ is parameterized by the population vector $\bm{x}$. 

During the fitting process, we model the intrinsic extinction ($A_{V,\,\star}$) using the foreground dust model and the Calzetti et al. (1994) extinction law with $R_{V}=4.05$ (Calzetti et al. 2000) for these emission-line dwarf galaxies. Furthermore, $A_{V,\,\star}$ is not constrained to be positive for reasons discussed in C05 and Mateus et al. (2006). Prior to the synthesis, we correct the calibrated spectra for redshift and Galactic extinction using the Cardelli et al. (1989) and O'Donnell (1994) Galactic extinction law with $R_{V}=3.1$, where the  $A_V$ values are taken from Schlegel et al.(1998) through the NASA/IPAC Extragalactic Database (NED). 

The observed spectra are normalized to the median flux between 4010 and 4060\r{A}, while the SSPs are normalized at $\lambda_0$=4020\r{A}. The continuum between 4730 and 4780\r{A}, which is generally free of emission lines, is used to measure the signal-to-noise ratio (S/N) of the observed spectra. During the fit, we mask the region of 20-30\r{A} around obvious emission lines for each object. We further give more weight to stellar absorption features (e.g., Ca\,{\sc ii}\,K\,$\lambda3934$, Ca\,{\sc ii} triplets) that are less affected by nearby emission lines and are among the strongest. For more details about the synthesis process, please refer to C05, Mateus et al. (2006) and Zhao et al. (2011).

In addition, we add a non-stellar component which represents the contribution of an AGN featureless continuum (FC) to the stellar base during the fitting process. We represent this non-stellar component by a $F_{\upsilon} \propto \upsilon^{-1.5}$ power law (e.g., C04;  Koski 1978; Riffel et al. 2009). The fractional contribution of this component to the total synthetic flux at $\lambda_0$=4020\r{A} is denoted by $x_{\rm {AGN}}$. As pointed out in C04, however, it is difficult for the STARLIGHT code to distinguish a $F_\nu \propto \nu^{-1.5}$ power law continuum from the spectrum of a dusty starburst. 
We postpone to \S3.7 a discussion on the possible biases in our results due to this degeneracy. We found that 9 galaxies in our sample present $A_{V,\,\star}<0$. For these objects, we re-ran the code by setting the lower limit $A_{V,\,\star}=0$. The fitted results do not change much except for three sources, which present luminosity-weighted ages lower than $10^9$~yr. Furthermore, it seems that negative extinction for passive galaxies is mainly caused by the limitation of the current base of SSPs (C05; Mateus et al. 2006). Therefore, the re-fitted results are adopted only for the three sources in the following analysis. In Figure~\ref{figexample}, we show a typical example of our fitting results for SDSS J133245.62+263449.3.

\begin{table*}[tb]
\begin{center}
\caption{Median values and standard deviations of the  Derived Stellar Population Properties
\label{fittedresult}}
\begin{tabular}{ccccccc}
      \hline
      \hline
   Sample& $\langle \log t_\star \rangle_L$ & $\langle \log t_ \star \rangle_M$ & $ \sigma_{L}(\log t_{\star})$ &$\sigma_{M}(\log t_{\star})$&$\langle Z_\star\rangle_L/Z_\odot$&$\langle Z_\star\rangle_M/Z_\odot$\\   
      \hline
      All &8.88$\pm$0.69&9.80$\pm$0.14&1.12$\pm$0.31&0.36$\pm$0.07&0.29$\pm$0.05&0.30$\pm$0.08\\
      \hline
      $f_{r}<20\%$ (S1)&8.58$\pm$0.73&9.81$\pm$0.15&1.17$\pm$0.25&0.40$\pm$0.12&0.27$\pm$0.07&0.26$\pm$0.15\\
      \hline
       $f_{r} \geq 20\%$ (S2) &8.94$\pm$0.62&9.80$\pm$0.13&1.06$\pm$0.33&0.36$\pm$0.06&0.29$\pm$0.05&0.32$\pm$0.07\\
       \hline
        $f_{r}<f_{r,\,{\rm med}}$\tablenotemark{$a$}&8.80$\pm$0.66&9.81$\pm$0.13&1.14$\pm$0.27&0.37$\pm$0.09&0.27$\pm$0.06&0.30$\pm$0.09\\
      \hline
       $f_{r} \geq f_{r,\,{\rm med}}$\tablenotemark{$a$} &8.91$\pm$0.67&9.79 $\pm$0.13&1.06$\pm$0.33&0.36$\pm$0.07&0.29$\pm$0.05&0.31$\pm$0.07\\
       \hline
      $x_{\rm AGN}=0$ (S3)&8.89$\pm$0.70&9.80$\pm$0.12&1.06$\pm$0.28&0.37$\pm$0.07&0.28$\pm$0.05&0.29$\pm$0.07\\
      \hline
       $x_{\rm AGN}>0$ (S4)&8.87$\pm$0.61&9.80$\pm$0.15&1.15$\pm$0.33&0.35$\pm$0.09&0.29$\pm$0.05&0.33$\pm$0.07\\      
      \hline
      \end{tabular}
      
      $^a$ $f_{r,\,{\rm med}}(\equiv 36.3\%)$ is the median value of $f_r$ for our sample.
      \end{center}
\end{table*}

\section{RESULTS and ANALYSIS}
Based on the the fitting results of STARLIGHT, we  obtain/derive the following parameters for the AGN host dwarf galaxies: mean stellar age,  mean stellar metallicity and the contribution of flux and mass from different SSPs, which can provide us very useful probes for the SFH studies. Both light- and mass-weighted mean stellar ages  and metallicities are estimated. 

\subsection{Mean Stellar Age}
As in C05, we adopt the following two equations to calculate the light- and mass-weighted mean stellar ages for the AGN host dwarf sample, i.e.
  \begin{center}
 \begin{equation}
      \langle \log t_{\star}\rangle_{L}=\sum\limits_{j=1}^{N_{\star}} x_{j}\log t_{j},
 \end{equation}
 \end{center}
for the light-weighted mean stellar age, and 
    \begin{center}
 \begin{equation}
      \langle \log t_{\star}\rangle_{M}=\sum\limits_{j=1}^{N_{\star}} \mu_{j}\log t_{j},
 \end{equation}
 \end{center}
for the mass-weighted mean stellar age. Here $x_{j}$ and $\mu_{j}$ represent the fractional contributions to the stellar luminosity and mass of the SSP with age $t_{j}$, respectively. $N_{\star}$ is the total number of SSPs. For a given object, $\langle t_{\star}\rangle_{L}$ is strongly affected by the recent SFH, while $\langle t_{\star}\rangle_{M}$ is mainly determined by the mass assembly history. As shown in C05, the uncertainties of $\langle \log t_{\star}\rangle_{M}$ and $\langle \log t_{\star}\rangle_{L}$ depend on the S/N of the input spectra.  Generally, the rms of $\langle \log t_{\star}\rangle_{L}$ is $<$0.1 and $\sim$0.15 dex respectively for ${\rm S/N} > 10$ and S/N$\sim$5. For $\langle \log t_{\star}\rangle_{M}$, the rms is $\sim$0.1 and $\sim$0.2 dex respectively for ${\rm S/N} > 10$ and ${\rm S/N} < 10$. For our AGN-host sample, there are eight sources having ${\rm S/N} < 5$, and the median value of S/N is 16.7. In Table \ref{fittedresult} we summarise the fitting results of these two kinds of ages for our (sub)samples.

\begin{figure*}[t]
\centering
\includegraphics[width=0.8\textwidth,bb=2 10 592 582]{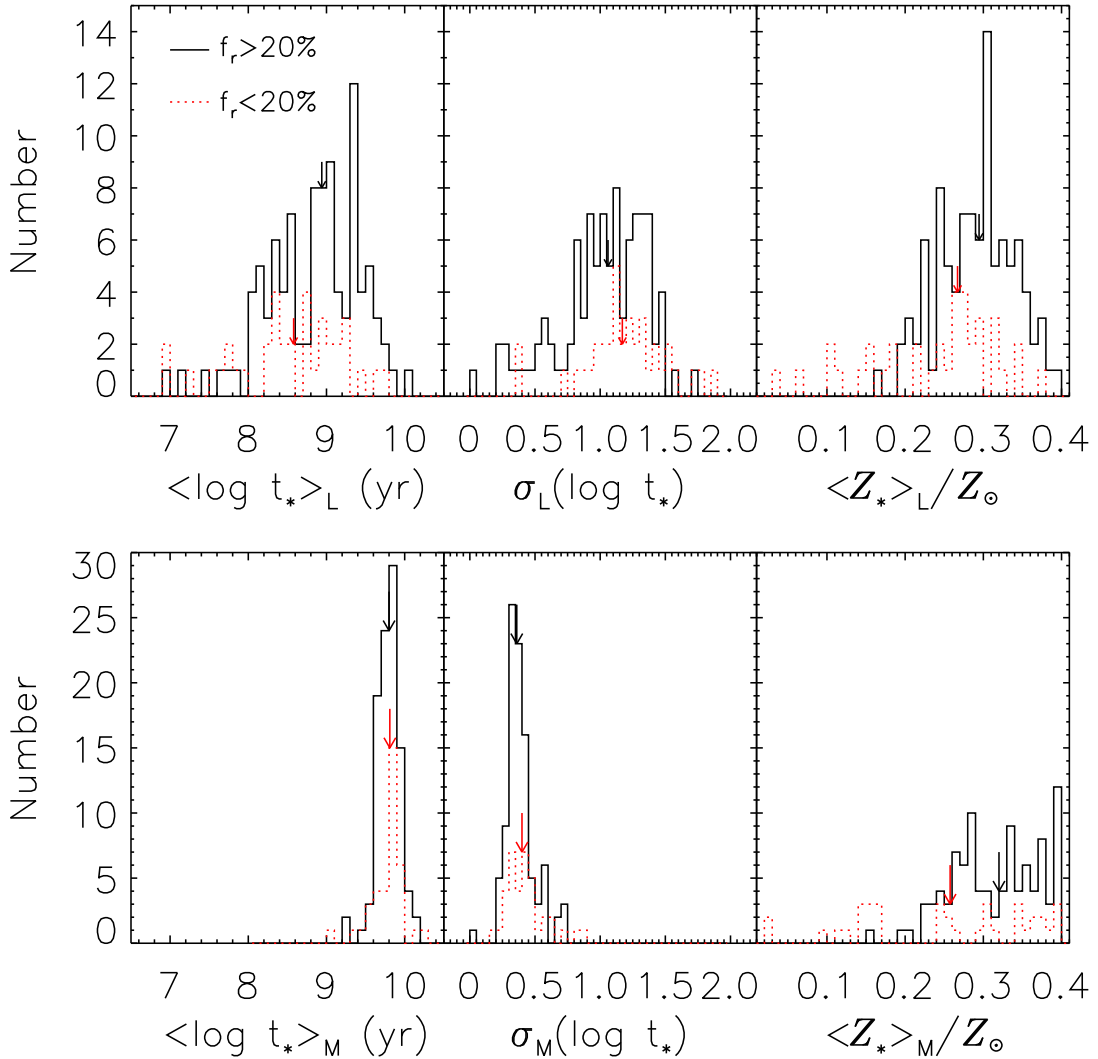}
\caption{The number distributions of the AGN-host dwarf sample in $\langle \log t_{\star}\rangle$ ({\it Left}), $\sigma(\log t_{\star})$ ({\it Middle}) and $\langle Z_{\star}\rangle$ ({\it Left}). Top and Bottom panels show the luminosity- and mass-weighted results, respectively. Arrows in each panel give the median values.}
\label{figage}
\end{figure*}

In the left column of Figure \ref{figage}, we show the distributions of $\langle \log t_{\star}\rangle_{L}$ and $\langle \log t_{\star}\rangle_{M}$ for the AGN sample. As shown in the top left panel of Figure \ref{figage}, the AGN sample has a wide distribution in $\langle \log t_{\star}\rangle_{L}$, ranging from $\sim$10~Myr to $\sim$10~Gyr, indicating a diversity of their SFHs. Out of these 136 sources, 15 galaxies (11\%) have $\langle \log t_{\star}\rangle_{L} < 8$, 68 (50\%) galaxies have $8< \langle \log t_{\star}\rangle_{L} < 9$  and 53 (39\%) galaxies have $ \langle \log t_{\star}\rangle_{L} >9$. This indicates that,  for these BPT-selected AGN-host dwarf galaxies, only $\sim$ten percent of sources have significant star-forming activities ($\langle t_{\star}\rangle_{L} < 100$~Myr), and 40 percent of systems are almost quiescent in the last 1~Gyr.

As shown in Table \ref{fittedresult}, the median values of $\langle \log t_{\star}\rangle_{L}$ are 8.9, 8.6 and 8.9, for the whole sample, subsample S1 and subsample S2, respectively. These median $\langle \log t_{\star}\rangle_{L}$ are much larger than that ($\sim$7.4) of the blue compact dwarf galaxies (BCDs; Zhao et al. 2011), which are a kind of starbursting dwarfs. Therefore, in general it is lack of intense star-forming activities in these AGN host dwarf galaxies (at least in the central region), which is consistent with the findings that the BPT method causes significant bias against AGN selection in low-mass, star-forming galaxies (Trump et al. 2015). 

Interestingly, our results are similar to those for a sample of 65 Seyfert 2 galaxies in C04, who used a similar method of analysis. From their Figure 18 we can see that the fraction of galaxies with $\langle \log t_{\star}\rangle_{L}<8$ and the median value of $\langle \log t_{\star}\rangle_{L}$ for those Seyfert 2s are almost identical to ours. Since the Seyfert 2s in C04 were identified with various methods, it seems difficult to explain why these two samples have similar results. One possible explain is that massive galaxies tend to have redder color and less specific star formation rate (star formation rate per stellar mass), and thus the combination of this with the selection bias of the BPT method causes the above coincidence. It needs more data, such as X-ray selected AGN host dwarf galaxies, to reach a solid conclusion.

Regarding $\langle \log t_{\star}\rangle_{M}$, both subsamples have a much narrower distribution, ranging from $\sim$1~Gyr to $\sim$10~Gyr, and all galaxies have $\langle \log t_{\star}\rangle_{M} > 1$~Gyr. Furthermore, subsample S1 has the same median value as subsample S2. Therefore, the stellar mass is contributed mostly by the old populations, similar to other types of emission-line dwarf galaxies (such as BCDs; Zhao et al. 2011).

 As shown in C05, the SFH of an object can be investigated in more detail using the two higher moments of the age distribution, i.e., the light- and mass-weighted standard deviations of the log age, namely,
  \begin{center}
 \begin{equation}
      \sigma_{L}(\log t_{\star})=\left[ \sum\limits_{j=1}^{N_{\star}} x_{j}(\log t_{j}-\langle \log t_{\star}\rangle_{L})^2\right ]^{1/2}
 \end{equation}
 \end{center}
 and
  \begin{center}
 \begin{equation}
      \sigma_{M}(\log t_{\star})=\left [ \sum\limits_{j=1}^{N_{\star}} \mu_{j}(\log t_{j}-\langle \log t_{\star}\rangle_{M})^2\right ] ^{1/2}.
 \end{equation}
 \end{center}
 Based on the definitions, it is easy to understand that $\sigma_{L}(\log t_{\star})$ and $\sigma_{M}(\log t_{\star})$ could discriminate between systems dominated by a single population and those which had bursty or continuous SFHs.

The middle panels of Figure \ref{figage} show the number frequency histograms of $\sigma_{L}(\log t_{\star})$ (top) and $\sigma_{M}(\log t_{\star})$ (bottom). On the one hand, irrespective of galaxy type and $f_r$, the median values (also see Table \ref{fittedresult}) of $\sigma_{L}(\log t_{\star})$ and $\sigma_{M}(\log t_{\star})$ are both larger than zero. This suggests that our sample galaxies could not be dominated by a single population. On the other hand, there exists a difference between $\sigma_{L}(\log t_{\star})$ and $\sigma_{M}(\log t_{\star})$ for both S1 and S2 subsamples: the median value of $\sigma_{L}(\log t_{\star})$ is in between 1.0 and 1.2, and the median $\sigma_{M}(\log t_{\star})$ is $\sim$0.4. This small $\sigma_{M}(\log t_{\star})$ indicates that these AGN-host dwarf galaxies have assembled the majority of stellar mass in a narrow time range, while the significantly larger $\sigma_{L}(\log t_{\star})$ suggests that most AGN host-dwarf galaxies have experienced repeated/continuous star formation activities.

 \subsection{Mean Stellar Metallicity}

\begin{figure}[t]
\centering
\includegraphics[angle=0,bb=22 40 440 585,width=0.47\textwidth]{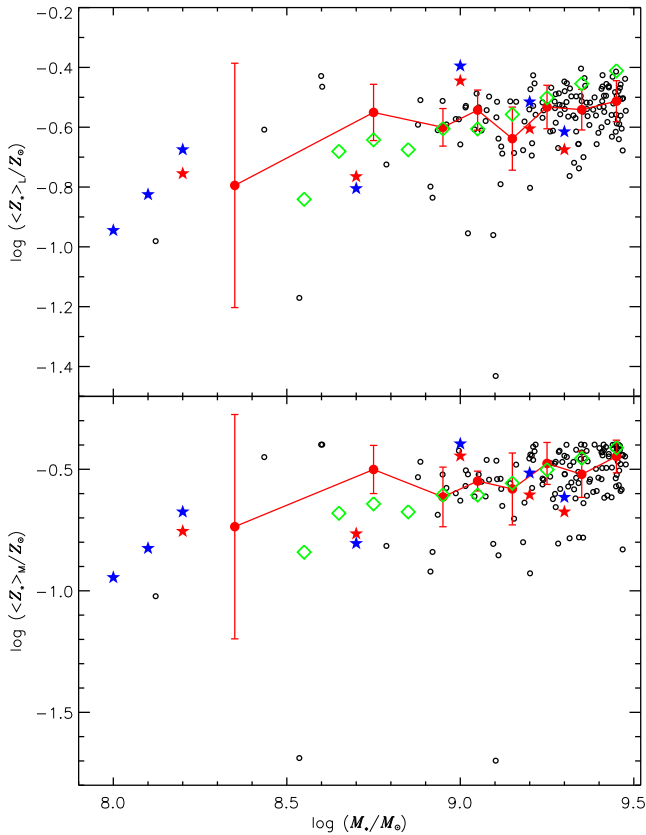}
\caption{Stellar mass versus light- ({\it Top}) and mass-weighted ({\it Bottom}) stellar metallicities for AGN-host dwarfs (black open circles). Red solid circles show the median values of each mass bin, with error bars giving the dispersion. For comparison, we also plot the star-forming galaxies from Zahid et al. (2017; green open diamonds), for which the stellar metallicities are obtained by fitting stacked spectra with stellar population synthesis models, and nearby galaxies (blue/red stars) with stellar metallicities measured with spectroscopy of individual (blue/red) supergiant stars (Davies et al. 2017 and references therein). Metallicities from Zahid et al. (2017) and Davies et al. (2017) have been rescaled to the solar units adopted in the current work ($Z_\odot=0.02$).}
\label{massz}
\end{figure}

  Similar to mean stellar ages, the light- and mass-weighted mean stellar metallicities are defined as  
  \begin{center}
 \begin{equation}
      \langle Z_{\star}\rangle_{L}=\sum\limits_{j=1}^{N_{\star}} x_{j}Z_{j}
 \end{equation}
 \end{center}
and
  \begin{center}
 \begin{equation}
 \langle Z_{\star}\rangle_{M}=\sum\limits_{j=1}^{N_{\star}} \mu_{j}Z_{j}, 
  \end{equation}
 \end{center}
 respectively (C05). The stellar metallicity ($\langle Z_\star\rangle$) derived here is specially useful for the study on the chemical properties of our AGN-host dwarfs, for which the nebular metallicity is very difficult (or even impossible) to be determined due to the  presence of an AGN. Nevertheless, there exist systematic biases in the derived $\langle Z_\star\rangle$ and $\langle \log t_{\star}\rangle$ at a level of $0.1-0.2$ dex due to the age-metallicity degeneracy (C05), resulting in part of the variations of $\langle \log t_{\star}\rangle$ presented in \S3.1.

In the right panels of Figure \ref{figage} we display the number distributions of $\langle Z_{\star}\rangle_{L}$ (top) and $\langle Z_{\star}\rangle_{M}$ (bottom) in units of solar metallicity ($Z_\odot=0.02$, following C05), for the two subsamples S1 (red dotted line) and S2 (black solid line). It can be seen from the figure that both $\langle Z_{\star}\rangle_{L}$ and $\langle Z_{\star}\rangle_{M}$ are in the range of $0.02-0.4\,Z_\odot$, and subsample S1 has a slightly larger dispersion and smaller median value of $\langle Z_\star\rangle$ (but consistent with each other within  the uncertainties; see Table \ref{fittedresult}). For the whole sample, both $\langle Z_{\star}\rangle_{L}$ and $\langle Z_{\star}\rangle_{M}$ have a median value of $\sim$0.3\,$Z_\odot$.

It is well known that stellar mass correlates with stellar metallicity or gas metallicity,  forming the so-called mass-metallicity relation (hereafter MZR; e.g., see Maiolino \& Mannucci 2019 for a review). Therefore, we can explore the AGN effects on the chemical evolution of the host galaxy by checking whether there is a MZR in these AGN-host dwarfs, and if exists, whether it follows the relations established in normal galaxies (e.g., Zahid et al. 2017 and references therein). To this purpose, we plot $M_\star$ against $\langle Z_{\star}\rangle$ in Figure \ref{massz}. It can be seen from the figure that the metallicities of these AGN-host dwarfs are well consistent with those of star-forming galaxies in Zahid et al. (2017) and nearby galaxies in Davies et al. (2017). Furthermore, the AGN-host dwarfs follow a similar MZR to star-forming galaxies. Our results suggest that AGNs unlikely have a strong impact on the chemical evolution of the host galaxy.

Figure \ref{massz} also demonstrates that a small number of our sources may suffer from the age-metallicity degeneracy, as they obviously deviate from the MZR in the plot. 
 As shown in the figure, most of these outliers have lower $\langle Z_{\star}\rangle$ comparing with other galaxies at similar $M_\star$. This could be associated with an overestimate of  $\langle \log t_{\star}\rangle$, especially $\langle \log t_{\star}\rangle_L$. According to C05, an underestimate of 0.2 dex in $\langle Z_{\star}\rangle_L$ would result in a comparable  overestimate in $\langle \log t_{\star}\rangle_L$. We can therefore estimate an upper limit to the bias in the derived mean stellar ages and metallicities due to the age-metallicity degeneracy to  $\sim 0.2-0.3$ dex.  For the two sources for which the metallicities could be underestimated by a factor of up to $\sim$10, their $\langle \log t_{\star}\rangle_L$ should be significantly overestimated. However, the uncertainties caused by the age-metallicity degeneracy will not affect our main results since only a small fraction of our sample galaxies suffer from this effect.

\subsection{Luminosity Fraction of the AGN}
\begin{figure}[t]
\centering
\includegraphics[angle=0,bb=60 68 270 243,width=0.47\textwidth]{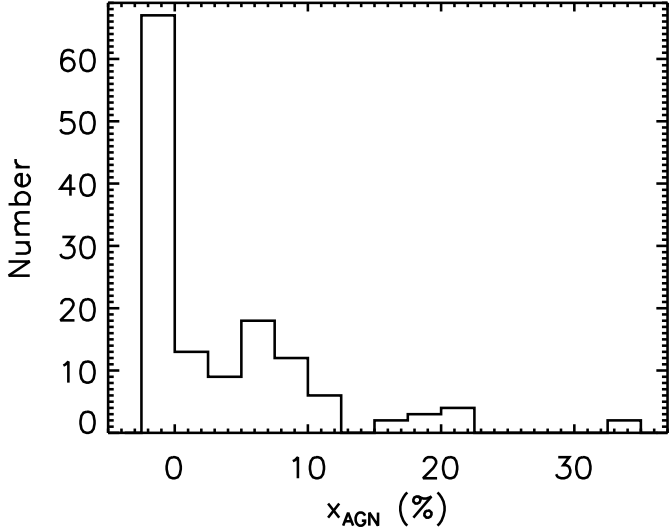}
\caption{The number distribution of $x_{\rm AGN}$ for AGN host dwarfs. The negative bin represents the number of galaxies which have $x_{\rm AGN}=0$.}
\label{figxagn}
\end{figure}

In Figure \ref{figxagn}, we plot the number distribution of the light fraction contributed by the AGN, $x_{\rm AGN}$. About a half of the sample galaxies (67 sources) have $x_{\rm AGN}=0$, 63 sources ($\sim$46\%) have $0\%<x_{\rm AGN}<20\%$, and only 6 sources ($\sim$4\%) have $x_{\rm AGN}>20\%$. For the whole sample, the mean $x_{\rm AGN}$ is $(4.2\pm6.6)\%$ \footnote{Throughout the paper, the uncertainty of a mean value is estimated using the standard deviation of the sample; the uncertainty of a median value is estimated using $1.48\times{\rm MAD}$ assuming normally distributed noise, where MAD is the median value of the absolute deviations from the median data.}, which is not surprising since the broad component, a direct evidence for the presence of an AGN continuum, are detected in H$\alpha$ in only 10 out of 136 sources (R13). For these 10 broad-line galaxies, indeed, the averaged $x_{\rm AGN}$ of $(11.4\pm7.5)\%$ is $>$3 times larger than that ($(3.7\pm6.2)\%$) for the rest sources. Our results confirm the prediction from Cid Fernandes \& Terlevich (1992), who demonstrated that the broad component in H$\alpha$ should become discernible when the scattered FC contributes $\sim$$>10\%$ to the optical continuum.

However, the broad components of H$\beta$ and H$\alpha$ lines are even not detected in four sources with $x_{\rm AGN}>20\%$. Thus we checked the spectra and the fitted results of these four sources. We found that 3 out of 4 sources have ${\rm S/N}<11$, and thus the non-detection of the broad component in H$\alpha$ might be attributed to their low S/N spectra. For the last source J104119.80$+$143641.1, it has a S/N$\sim$20, and the fitted results seem reasonable. After carefully inspecting its spectrum, we identified a weak broad component in H$\alpha$ in this object.

For a given source, $x_{\rm AGN}$ is a function of the aperture size $r_{\rm A}$, namely the larger $r_{\rm A}$, the smaller $x_{\rm AGN}$. Since the physical size covered by the 3$\arcsec$-fiber is different for our sample galaxies, we would like to check to what extent this will affect our fitted $x_{\rm AGN}$. In Figure \ref{figfrxagn}, $x_{\rm AGN}$ is plotted against the light fraction $f_r$. There is no correlation between these two parameters, suggesting that the fitted $x_{\rm AGN}$ would not be mainly determined by the aperture size for our sample galaxies.   

\begin{figure}[t]
\centering
\includegraphics[angle=0,bb=60 68 270 243,width=0.47\textwidth]{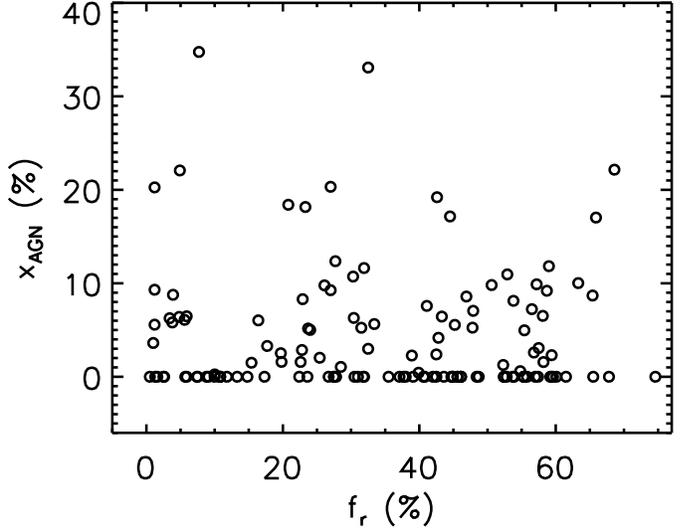}
\caption{Light fraction ($f_r$) within the SDSS 3\arcsec-aperture plotted against $x_{\rm AGN}$.}
\label{figfrxagn}
\end{figure}

\begin{figure*}[t!]
\begin{center}
\includegraphics[angle=0,bb=60 50 570 456, width=0.85\textwidth]{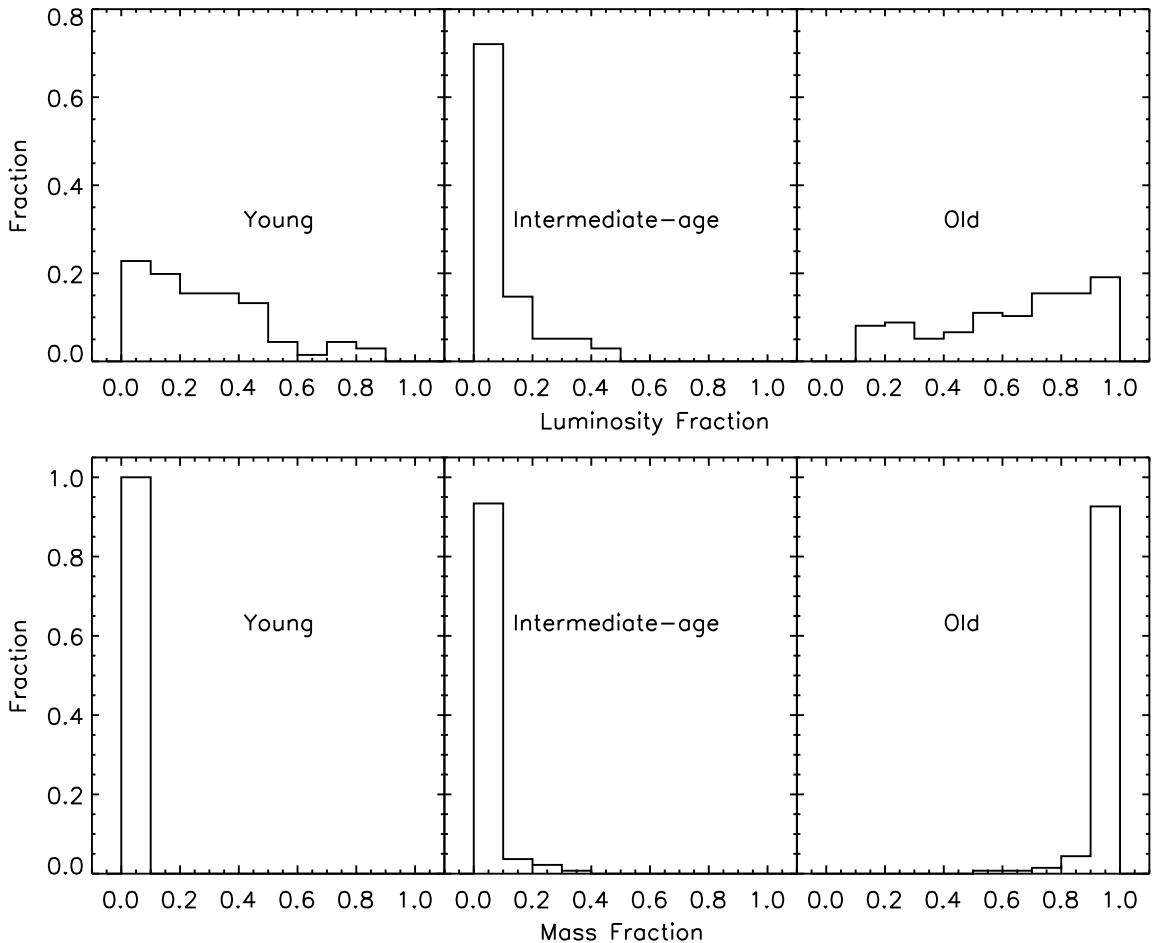}
\end{center}
\caption{Number distributions of the luminosity (Top) and mass (Bottom) fractions for the young ($t< 10^8$ yr), intermediate ($10^8 < t < 10^9$ yr), and old ($t > 10^9$ yr) stellar populations.}
\label{figyio}
\end{figure*}

\subsection{The Approximate SFHs}
As discussed in C05, there exist large uncertainties in the  individual components of $\bm{x}$ ($\bm{\mu}$), while the binned vectors of $\bm{x}$ ($\bm{\mu}$), i.e. the young ($t< 10^8$yr), intermediate ($10^8 < t < 10^9$ yr), and old ($t > 10^9$ yr) populations, have uncertainties less than 0.05, 0.1, and 0.1, respectively, for ${\rm S/N} >10$. In order to have a robust (though coarse) SFH of each galaxy, here we binned $\bm{x}$ into these three components ($x_{\rm Y}$ ($\mu_{\rm Y}$), $x_{\rm I}$ ($\mu_{\rm I}$) and  $x_{\rm O}$ ($\mu_{\rm O}$), respectively). Since $\bm{x}$ given by the stellar synthesis usually do not add up to 100\%, and a light fraction is associated to the AGN, the final luminosity fractions were normalised to have $x_{\rm Y}+x_{\rm I}+x_{\rm O}=1$.

 As shown in the top panels of Figure~\ref{figyio}, the old population contributes most light ($>50\%$) for the majority ($\sim$70\%) of our sample galaxies (mean $x_{\rm O}$ of $(64.2\pm26.5)\%$), although the young population also appears in significant (but widely varying) proportions (mean $x_{\rm Y}$ of ($28.2\pm21.5)\%$). There are only 18 ($\sim$13\%) sources having more than 50\% of their light contributed by the young population. Furthermore, the intermediate-age population contributes $<$50\% light for all objects, and contributes $<$20\% light for the vast majority ($\sim$90\%). The relatively small fraction of the young population and the lack of the intermediate-age population (mean $x_{\rm I}=(7.6\pm12.0)\%$) suggests that, in general the star formation in these galaxies is likely repeated at a mild level with long quiescent periods. Compared with our results, the Seyfert 2 sample in C04 has less light from the young and old populations, but more light from the intermediate-age population, with the mean fractions of $x_{\rm Y}=(14.0\pm15.6$)\%, $x_{\rm I}=(32.4\pm32.2$)\% and $x_{\rm O}=(53.6\pm33.0$)\%. 

The bottom panels of Figure~\ref{figyio} display the distributions of mass fraction from the three populations. Clearly, the young population only has a tiny contribution ($<$10\%) to the stellar mass, with a mean fraction of $\mu_{\rm Y}=(0.8\pm1.4)\%$, while the old population contributes an overwhelming majority (mean $\mu_{\rm O}=(96.9\pm6.3)\%$) of stellar mass in these objects. Similar to the young population, the intermediate-age population also only contributes a very small fraction, with the mean $\mu_{\rm I}$ of $(2.2\pm5.4)\%$, though its contribution can be up to $\sim$40\% in some objects. 

 Comparing our results with those of galaxies with $10^9<M_\star<3\times10^{9}\,M_\odot$ (122 sources in our sample are in this mass range) in Kauffmann (2014), we find that the young population in AGN-host dwarfs contributes less fraction to the present-day stellar mass. For example, the fraction of galaxies with burst mass fraction ($F_{\rm burst}$, i.e., mass formed during the starburst starting 100 Myr ago divided by the total stellar mass) larger than 5\%  is in the range of $0.25-0.4$, while only 5 out of 122 sources (e.g., 0.04) have $\mu_{\rm Y}>5\%$ in our sample. Furthermore, the mass fraction of the old population in AGN-host dwarfs is closer to that of dwarf transitions and dSph/dE (Weisz et al. 2011a,b). Therefore, AGNs in dwarfs seem to suppress the current star formation of the host galaxy. However, our sample may suffer from uncertainties caused by selection bias and/or small size of the sample used. In the future we will further explore the AGN effects by using samples selected with different methods, and by comparing with carefully selected control samples.

\subsection{Relation with the Host Galaxy Morphology}
\begin{figure}[t]
\centering
\includegraphics[angle=0,bb=30 41 440 583,width=0.47\textwidth]{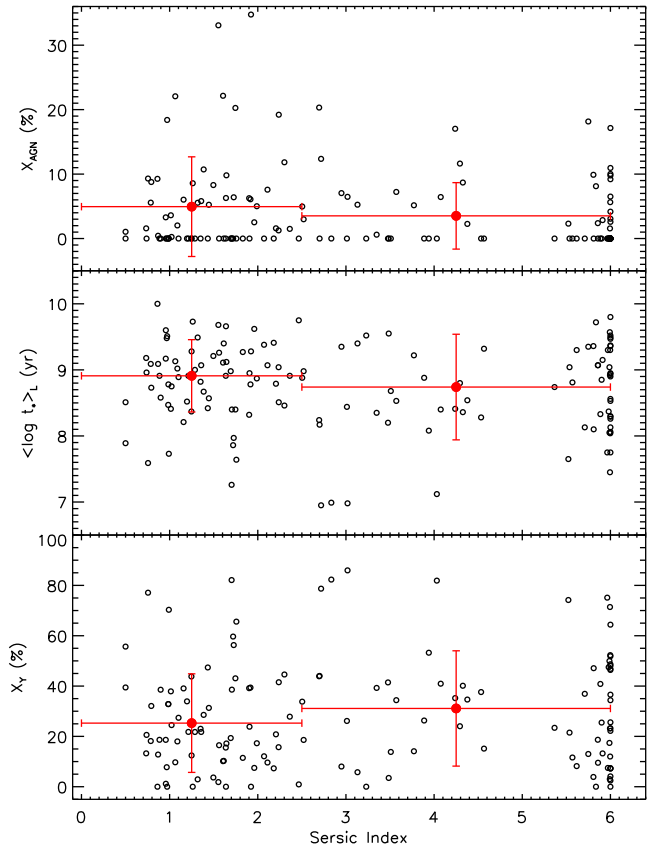}
\caption{Plot of Sersic index $n$ of the host galaxy versus  light fraction of the AGN (Top), light-weighted stellar age (Middle) and luminosity fraction of the young stellar population (Bottom). The red solid circles represent the median (mean) values of $\langle \log t_{\star}\rangle_{L}$ ($x_{\rm Y}$ and $x_{\rm AGN}$) for early-type ($n>2.5)$ and late-type ($n<2.5$) sources,  with error bars in the $y$-axis giving the dispersion of the data within each bin.}
\label{indexage}
\end{figure}

It is known that generally the SFH of a galaxy is related to its morphology, i.e. late type galaxies have bluer colour and higher star formation rate than early types (e.g., Kennicutt 1998). For more massive AGN hosts (e.g., Seyfert 2s), however, contradictory results are presented by different works: Storchi-Bergmann et al. (2001) found that the fraction of galaxies with recent (circum-)nuclear starburst increases along the Hubble sequence for a sample of 35 objects, while C04 and Morelli et al. (2013) found no correlation between star formation and morphology of the host galaxy for samples respectively consisting of 65 and 20 sources. A very recent work by Mallmann et al. (2018) also found no significant difference between SFHs of early- and late-type host galaxies, based on the spatially resolved stellar populations for a sample of 62 AGNs selected from the SDSS-IV MaNGA survey (Bundy et al. 2015; Law et al. 2015; Blanton et al. 2017). 

To investigate whether there exists a correlation between star formation properties and morphology for our dwarf sample, we plot the S\'{e}rsic index $n$, which  is adopted from the NSA, against $\langle \log t_{\star}\rangle_{L}$ and $x_{\rm Y}$ in Figure \ref{indexage}, with the median (mean) value of $\langle \log t_{\star}\rangle_{L}$ ($x_{\rm Y}$) overlaid. Here $n$ was obtained by fitting the galaxy's two-dimensional surface brightness distribution with a S\'{e}rsic function, and generally a larger $n$ indicates an earlier type.  We further adopt $n=2.5$ as the dividing line between early- ($n>2.5$) and late-type ($n<2.5$) galaxies according to Barden et al. (2005), which has discriminated between visually classified early- and late-type objects.

From the figure we can see that both $\langle \log t_{\star}\rangle_{L}$ and $x_{\rm Y}$ have no correlation with $n$, indicating no dependence of the (circum-)nuclear SFHs on the host morphology. Our result is consistent with those for Seyfert 2s (e.g., C04; Morelli et al. 2013; Mallmann et al. 2018). For our sample galaxies, however, the S\'{e}rsic index $n$ from a single-component fit might not accurately represent the host galaxy morphology if a central bright point source  is present. To further check this, we also plot $x_{\rm AGN}$ versus $n$ in the top panel of Figure \ref{indexage}, and find no correlation between these two parameters. Therefore, the (circum-)nuclear SFHs of these AGN host dwarfs are truly independent on its morphology.  However, more detailed measurements of the host morphologies with multi-component fits to deeper images are needed before a robust conclusion can be drawn.

\subsection{Relation with the \OIIII\ Luminosity }

\begin{table}[t]
\begin{center}
\caption{Summary of the Spearman's Rank Correlation Coefficients}
\label{spearmancc}
\begin{tabular}{lccccc}
\hline
\hline
\scriptsize
Relation&Sample\tablenotemark{$a$}&Subsample&No.&$\rho$&$p$-value\\
\hline
\multirow{4}{*}{[O\,{\sc iii}]-$\langle t_{\star}\rangle_{L}$} & All  & ---&136&$-0.20$&$1.8\times10^{-2}$\\
&$L$$>$10$^{39}$&---&91&$-0.41$&$6.4\times10^{-5}$\\
&&AGN&33&$-0.34$&$4.9\times10^{-2}$\\
&&Composite&58&$-0.52$&$2.6\times10^{-5}$\\
\hline
\multirow{4}{*}{[O\,{\sc iii}]-$x_{\rm Y}$} & All  & ---&136&0.19&$2.6\times10^{-2}$\\
&$L$$>$10$^{39}$&---&91&0.42&$2.8\times10^{-5}$\\
&&AGN&33&0.37&$3.6\times10^{-2}$\\
&&Composite&58&0.55&$7.5\times10^{-5}$\\
\hline
\end{tabular}
\end{center}
$^a$ $L$$>$10$^{39}$ refers to the sources having $L_{\rm [O\,{\scriptsize \textsc{iii}}]} > 10^{39}$~erg~s$^{-1}$.
\end{table}

\begin{figure}[t]
\centering
\includegraphics[angle=0,bb=6 62 466 550, width=0.47\textwidth]{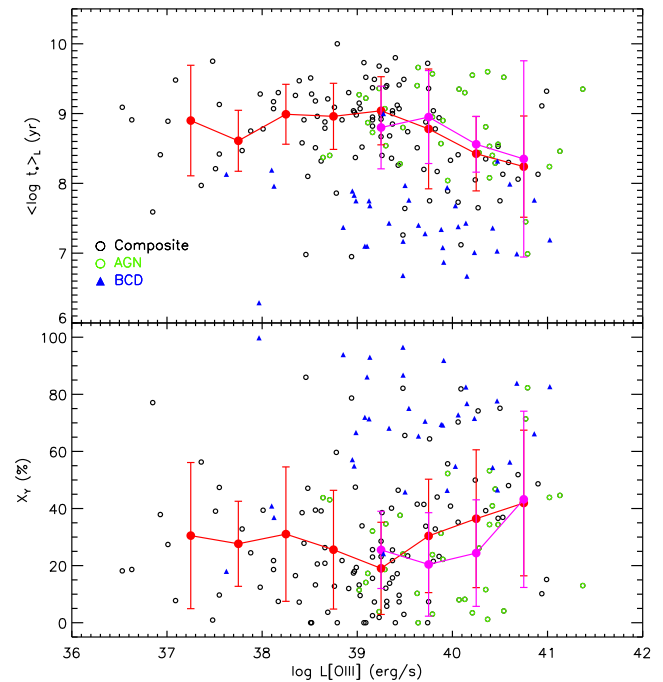}
\caption{Plot of the \OIII\ luminosity versus light-weighted stellar age (Top) and luminosity fraction of the young stellar population (Bottom). The red and purple solid circles represent the median (mean for $x_{\rm Y}$) values in each \LOIII\ bin for the whole sample and the 35 pure AGNs (green open circles), respectively.  The error bars in the $y$-axis show the dispersion of the data within each bin. For comparison, BCDs from Zhao et al. (2011) are also plotted as blue triangles.}
\label{AGNage12}
\end{figure}

AGN activities are found to be connected with the star formation properties of their host galaxies (e.g., Lutz et al. 2010; Rovilos et al. 2012; Santini et al. 2012; Hickox et al. 2014; Esquej et al. 2014). It has been shown in Kauffmann et al. (2003) that the \OIII\ line luminosity, \LOIII, is a good tracer of AGN activity (see also Trump et al. 2015). This is especially true for metal-rich galaxies, since \OIII\ is relatively weak in these objects in the absence of an optical AGN. Furthermore, AGN dominates the [O\,{\sc iii}] line emission (Kauffmann et al. 2003; Trump et al. 2015) in sources with $L_{\rm [O\,{\scriptsize \textsc{iii}}]} > 10^{7.5}\,L_\odot$ ($\sim$10$^{41}$~erg~s$^{-1}$). Therefore, it is intriguing to check whether the SFHs of host dwarfs have a correlation with the \OIII\ line luminosity.

As shown in Figure~\ref{AGNage12}, the extinction-corrected (see \S3.7) \LOIII\ of our sample galaxies is in the range of $10^{36.5}-10^{41.5}$~erg~s$^{-1}$ (i.e. $10^3-10^8\,L_\odot$), and 12 objects can be characterized as ``strong AGN" as their $L_{\rm [O\,{\textsc {\scriptsize {iii}}}]]} \geq 10^7\,L_\odot$ (e.g., Kauffman et al. 2003). Pure AGNs (green circles in the figure) tend to have higher \LOIII, with a median value of $\sim$10$^{40.0}$~erg~s$^{-1}$, than the composite objects (median \LOIII\ of $\sim$10$^{39.2}$~erg~s$^{-1}$). 

In Figure \ref{AGNage12} we plot $\langle \log t_{\star}\rangle_{L}$ (top panel) and $x_{\rm Y}$ (bottom panel) as a function of \LOIII, and find no obvious correlations. The Spearman correlation coefficients ($\rho$) are $-0.20$ and $0.19$ (see Table \ref{spearmancc}), respectively, with $p$-values of $1.8\times10^{-2}$ and  $2.6\times10^{-2}$. However, it seems that there exists a mild (anti-) correlation between $x_{\rm Y}$ ($\langle \log t_{\star}\rangle_{L}$) and \LOIII, for sources with $L_{\rm [O\,{\scriptsize \textsc{iii}}]}$$>$$10^{39}$~erg~s$^{-1}$, with $\rho$ of $0.42$ ($-0.41$), and $p$-value of $\ll 0.001$.

Rembold et al. (2017) studied the nuclear stellar populations of 62 AGNs in more massive galaxies with $10^{9.4}<M_\star<10^{11.5}\,M_\odot$, which have \LOIII\ in the range of $10^{39}-10^{42}$ erg s$^{-1}$. They argued that objects with higher \LOIII\ are younger than the control sample, while the low-luminosity AGNs (i.e. smaller \LOIII) are older. For the sources with $L_{\rm [O\,\scriptsize {\textsc {iii}}]} > 10^{39}$~erg~s$^{-1}$, it seems that our results are consistent with the findings in Rembold et al. (2017), though their sample don't include any composite objects, in which the AGN contribution to \LOIII\ is found to be in the range of 50\% to 90\% (but for more massive sources; Heckman et al. 2004). We further check the 33 pure AGNs in our sample, and find the (anti-) correlations still exist, and have a similar $\rho$  but a larger $p$-value. 

From Table \ref{spearmancc} we note, however, the composite systems with $L_{\rm [O\,{\scriptsize \textsc {iii}}]}$$> 10^{39}$~erg~s$^{-1}$ show the strongest correlation. Therefore, the increasing $x_{\rm Y}$ (decreasing $\langle \log t_{\star}\rangle_{L}$) with increasing \LOIII\ might be simply due to the increment of young massive stars which is not associated to the AGN phenomenon. To further explore this question, in Figure \ref{AGNage12} we also plot the results for BCDs, which are starburst systems (Zhao et al. 2011). From the figure, we can see that (1) There is no correlation between \LOIII\ and $\langle \log t_{\star}\rangle_{L}$ (or $x_{\rm Y}$) for BCDs; (2) In general BCDs are much younger than AGN-host dwarfs at a given \LOIII, and thus the \OIII\ emission in most composites should be dominated by AGN but not star formation. Hence, the apparently stronger correlation in composites seems only statistical since the composites have a relatively larger sample than the pure AGNs.

Our results suggest that there should be some physical connections between the nuclear star-forming and AGN activities. For example, both BH accretion and star formation could be fueled by the same gas reservoir (e.g., Trump et al. 2015), in the sense that nuclear activities are stronger when more gas is present.} 

\subsection{Internal Extinction}
 The internal extinction of a galaxy can be measured independently using the continuum and nebular lines. For the continuum, the stellar visual extinction, $A_{V,\,\star}$, is returned by STARLIGHT, modeled as due to a foreground dust screen. For the nebular lines, we make use of the reasonable assumption that the intrinsic Balmer line ratios are equal to Case B recombination, and the Calzetti et al. (1994) reddening law to derive the nebular extinctions from the observed H$\alpha$/H$\beta$ Balmer decrement, namely 
\[A_{V,\,{\rm neb}}=7.93 \times \log \left(\frac{F_{\rm H\alpha}/F_{\rm H\beta}}{I_{\rm H\alpha}/I_{\rm H\beta}}\right),\]
where $F_{\rm H\alpha}/F_{\rm H\beta}$ and $I_{\rm H\alpha}/I_{\rm H\beta}$ are the observed and intrinsic Balmer decrements, respectively. Here we adopt $I_{{\rm H}\alpha}/I_{{\rm H}\beta}=3.1$ appropriate for AGNs (e.g., Osterbrock \& Ferland 2006), for an electron temperature of 10$^4$~K and an electron density of 100~cm$^{-3}$.

For our sample galaxies, the median value of $\log (F_{\rm H\alpha}/F_{\rm H\beta})$ is $0.60\pm0.08$ (i.e. $A_{V,\,{\rm neb}}$$\sim$0.9~mag). For comparison, we also obtain the median $\log (F_{\rm H\alpha}/F_{\rm H\beta})=0.53\pm0.03$ (i.e., $A_{V,\,{\rm neb}}$$\sim$0.6~mag) of the star-forming dwarf galaxies for the stellar mass bin around $\log (M_\star/M_\odot)=9.3$ (the median mass of our sample) presented in Garn \& Best (2010; see their Figure 4), using $A({\rm H}\alpha)=6.54\times\log (F_{\rm H\alpha}/F_{\rm H\beta})-2.98$ and assuming $I_{{\rm H}\alpha}/I_{{\rm H}\beta}=2.86$. These results suggest that the AGN-host dwarfs seem to have a higher extinction, which could be naturally explained if photons from the AGN suffer dust obscuration both from the narrow line region (e.g., Lu et al. 2019) and the host galaxy, and dominate the observed line fluxes. 

Interestingly, we find that the median value of $A_{V,\,{\rm neb}}/A_{V,\,\star}$ ratios for the 120 AGN-host dwarfs, which have positive $A_{V,\,{\rm neb}}$ and $A_{V,\,\star}$, is $3.4\pm2.6$, $\sim$1.5 times larger than that  ($\equiv 1/(0.44\pm0.03)=2.3\pm0.2$) found in the detailed studies of nearby star-forming galaxies by Calzetti et al. (2000; see also Charlot \& Fall 2000), roughly consistent with that there is an additional source of dust obscuration in AGN-host dwarfs. Could this larger $A_{V,\,{\rm neb}}/A_{V,\,\star}$ ratio be caused by the degeneracy between the AGN FC and an dusty starburst as mentioned in \S2.2? This can be examined by comparing the $A_{V,\,{\rm neb}}/A_{V,\,\star}$ ratios of the two subsamples S3 and S4, which have median $A_{V,\,{\rm neb}}/A_{V,\,\star}$ of 3.40 and 3.35 respectively. Therefore, it seems that this degeneracy has negligible effect. However, we caution the reader that such differences are only tentative and prone to large uncertainties. A much larger sample is needed to confirm this tentative result.


\section{Discussion}
\subsection{Aperture Effects}
The biases caused by using the small apertures to measure the galaxy spectra  need to be considered, for the SDSS is a fiber-based survey. Fortunately, Kewley et al. (2005) have investigated this problem in detail and conclude that it is required 20\% of the galaxy light in order to minimize the aperture effects. For our sample, 99 out of 136 galaxies  present $f_r \geq 20\%$, with a median $f_r$ of $f_{r,\,{\rm med}}= 36.3\%$. It can be seen from Table \ref{fittedresult} that the differences of the fitted results between S1 ($f_r < 20\%$) and S2 ($f_r \geq 20\%$) are small, and the differences become even smaller when the two subsamples ($f_r<f_{r,\,{\rm med}}$ and $f_r \geq f_{r,\,{\rm med}}$) are compared. Therefore, the aperture effect does not seem  to severely affect our main results.

As shown by Zhao et al.(2011), the fitted $\langle \log t_{\star}\rangle_{L}$ has a good correlation with the $(g-r)$ color. Therefore, we can further check the aperture effect by comparing the fiber $(g-r)$ color with the whole galaxy for our AGN sample. We find that the 3$^{\prime\prime}$ region has almost identical $(g-r)$ color with the whole galaxy, with the median $\Delta (g-r)$($\equiv (g-r)_{\rm fiber}-(g-r)_{\rm whole})$ of $-0.01\pm0.07$~mag, which will result in a small difference in  $\langle \log t_{\star}\rangle_{L}$ according to Zhao et al. (2011).

\subsection{Effect of an AGN FC on the Derived SFHs}
As pointed out in C04 and discussed in \S3.2, STARLIGHT might improperly assign a dusty starburst to an AGN FC, resulting in an overestimation of the mean stellar age. Therefore, it is crucial to check the  effect of an AGN FC on the derived stellar population properties for our sample galaxies.  

As shown in Figure~\ref{figxagn}, about half our sample galaxies have $x_{\rm AGN} = 0$. However, these galaxies should have AGN contributions to the optical continuum though it could be intrinsically small. To have a rough estimation on how this biases our mean stellar age, we assume that we did not include an AGN FC  in the input library, and then compare our results with those from Cardoso et al. (2017), who have quantitatively examined the effect of an AGN FC on the estimation of SFHs of galaxies.       

As shown in Figures 7 and 8 in Cardoso et al. (2017), the neglect of the AGN FC will cause an overestimation of $\sim$$0.05-0.3$ dex (depending on the shape of the FC) for the light-weighted stellar age when $ \langle \log t_{\star}\rangle_{L}=7$, and an underestimation of $\sim$$0.4-0.6$ dex when $\langle \log t_{\star}\rangle_{L}=10$, for an intrinsic $x_{\rm AGN} = 0.2$. For $ \langle \log t_{\star}\rangle_{M}$, it will lead to an overestimation of $0.1-0.5$ dex when $\langle \log t_{\star}\rangle_{M}=9$ to $<$0.1 dex when $\langle \log t_{\star}\rangle_{M}=10$. Since these 67 galaxies have $ 7 < \langle \log t_{\star}\rangle_{L} < 10$ (median value of 8.90) and $ 9<\langle \log t_{\star}\rangle_{M} < 10$ (median value of 9.80), and generally their intrinsic $x_{\rm AGN}$ are far less than 0.2, there should only exist a negligible effect on the estimation of the mean stellar ages when the fitted $x_{\rm AGN}=0$. 

\begin{figure}[t]
\centering
\includegraphics[angle=0,bb=8 30 327 221,width=0.47\textwidth]{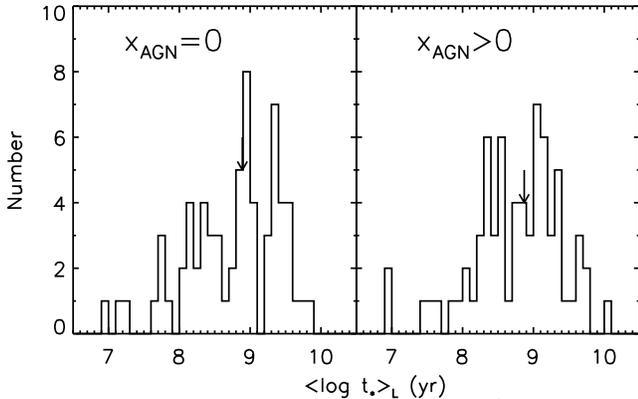}
\caption{The number distributions of $\langle \log t_{\star}\rangle_{L}$ for the two subsamples, $x_{\rm AGN} = 0$ ({\it left}) and $x_{\rm AGN} > 0$ ({\it right}).  The arrows give the median values.}
\label{agndis}
\end{figure}

This can be further checked by dividing the sample into two subsamples, $x_{\rm AGN}=0$ and $x_{\rm AGN}>0$ (S3 and S4 in Table \ref{fittedresult}, respectively). Figure \ref{agndis} shows the distributions of $\langle \log t_{\star}\rangle_{L}$ for S3 (left panel) and S4 (right panel). In general, both S3 and S4 have a similar range of $\langle \log t_{\star}\rangle_{L}$. The median $\langle \log t_{\star}\rangle_{L}$ are almost the same, 8.89 and 8.87 for S3 and S4, respectively. Both subsamples also have an identical median $\langle \log t_{\star}\rangle_{M}$ (9.80; Table \ref{fittedresult}). Furthermore, these two subsamples have a very similar median value of the extinction ratios ($A_{V,\,{\rm neb}}/A_{V,\,\star}$; see \S3.7), indicating that generally $A_{V,\,\star}$ for S4 is not underestimated. Therefore, these results confirm that the presence (or absence) of an AGN FC returned by STARLIGHT will not much affect the derived SFHs of our dwarf galaxies. 

\section{SUMMARY}
In this paper we present detailed results of our stellar population synthesis for a sample of 136 AGN-host dwarf galaxies selected by Reines et al. (2013) from SDSS DR8, using the  STARLIGHT code and fiber spectra obtained by SDSS. Our main results are:
\begin{enumerate}
     \item The optically selected AGN-host dwarfs show a diversity of SFHs. The old population ($>$10$^9$~yr) contributes most light for the majority of the sample galaxies, the young population ($<$10$^8$~yr) also appears in significant but widely varying fractions, while the intermediate-age population ($10^8<t<10^9$~yr) generally has little contribution to the optical continuum at 4020 \AA.  
     
     \item The light-weighted stellar ages are in the range of $10^7-10^{10}$~yr, with a median value of $10^{8.9}$~yr, and the mass-weighted stellar ages are in the range of $10^9-10^{10}$~yr, with a median value of $10^{9.8}$~yr. The light-weighted standard deviation of the log age is in the range of $0-2$, with a median value of $\sim$1.1, while the mass-weighted standard deviation of the log age has a much narrower range ($0-0.8$) and smaller median value ($\sim$0.4), indicating that the star-forming activities may occur repeatedly, but the stellar mass is contributed by old stars and has assembled within a short period of time.
   
      \item The AGN-host dwarf galaxies follow a similar mass-metallicity relation to normal star-forming galaxies, indicating that AGNs unlikely have a great effect on the chemical evolution of the host galaxies.

     \item Generally the AGN FC only contributes a small fraction to the optical continuum, with a mean value of $x_{\rm AGN}=(4.2\pm6.6)\%$, and about a half of the sample galaxies have no detectable AGN FC contribution. However, $x_{\rm AGN}$ is, on average, about 3 times higher in broad-line AGNs.

     \item There is no correlation between the derived SFHs and morphology of the host galaxy, agreeing with the results for Seyfert 2s in the literature.

     \item A mild (anti-)correlation between $x_{\rm Y}$ ($\langle \log t_{\star}\rangle_{L}$) and \LOIII\ is found, indicating that the central black hole and star formation may be fueled by the same gas reservoir.
     
     \item The nebular extinctions in these AGN-host galaxies are found to be larger than star-forming dwarfs with similar stellar masses, and we speculate that the photons from AGNs might be obscured by the dust in the narrow line region and the host galaxy.   
\end{enumerate}

\begin{acknowledgements}
We thank the anonymous  referee for his/her careful reading and thoughtful comments which improved the paper. The work is supported by the National Key R\&D Program of China grant No. 2017YFA0402704, the Natural Science Foundation of China (NSFC; grant Nos. 11991051, 11421303 and 11973039), and the CAS Pioneer Hundred
Talents Program. The STARLIGHT project is supported by the Brazilian agencies CNPq, CAPES, and FAPESP and by the France–Brazil CAPES/Cofecub program. All the authors acknowledge the work of the Sloan Digital Sky Survey (SDSS) team. Funding for SDSS-III has been provided by the Alfred P. Sloan Foundation, the Participating Institutions, the National Science Foundation, and the U.S. Department of Energy Office of Science. The SDSS-III web site is http://www.sdss3.org/. SDSS-III is managed by the Astrophysical Research Consortium for the Participating Institutions of the SDSS-III Collaboration including the University of Arizona, the Brazilian Participation Group, Brookhaven National Laboratory, Carnegie Mellon University, University of Florida, the French Participation Group, the German Participation Group, Harvard University, the Instituto de Astrofisica de Canarias, the Michigan State/Notre Dame/JINA Participation Group, Johns Hopkins University, Lawrence Berkeley National Laboratory, Max Planck Institute for Astrophysics, Max Planck Institute for Extraterrestrial Physics, New Mexico State University, New York University, Ohio State University, Pennsylvania State University, University of Portsmouth, Princeton University, the Spanish Participation Group, University of Tokyo, University of Utah, Vanderbilt University, University of Virginia, University of Washington, and Yale University.
\end{acknowledgements}

\end{document}